# EXPERIMENTAL CONFIRMATION OF THE GRAVITATION FORCE NEGATIVE TEMPERATURE DEPENDENCE

A. L. Dmitriev and E. M. Nikushchenko

*St. Petersburg State University of Information Technologies, Mechanics and Optics,
St. Petersburg, 49 Kronverksky prospect, 197101, Russia, Tel/Fax:+7(812)3154071,
alex@dmitriyev.ru*

**Abstract:** *The experiment with weighing PZT-piezoelectric ceramics, heated up by a high-frequency signal for the temperature of 1.6 $^0C$ is briefly described. The negative change of piezoelectric ceramics weight having relative value of $\gamma \approx -4.1 \cdot 10^{-6} K^{-1}$ is confidently registered. The sign and the order of the value of relative temperature change of piezoelectric ceramics weight correspond to the measurements of weight of non-magnetic metal bars which were conducted earlier. What is emphasized as expedient for development of physics of gravitation is conducting similar measurements with use of various materials as samples and in a wide range of temperatures.*

The problem of influence of bodies' temperature on the force of their gravitational interaction has been discussed since long ago and the first precision experiments in this field were already carried out at the beginning of the XXth century [1]. The decline of interest to such researches, which followed later, might be explained by the authority of the general theory of relativity according to which the temperature dependence of force of gravitation practically can not be observed [2]. The next stage of experimental studies of the said specified problem fell to the beginning of the current millenium when in Russia there were published the results of laboratory measurements of temperature dependence of weight of metal bars, indicating an appreciable negative temperature dependence of the gravitation forces [3-5]; recently these results were confirmed in works of Chinese scientists [6].

The physical substantiation of relatively strong influence of temperature on force of gravitation consists in deep interrelation of electromagnetic and gravitational interactions, and their dependence on the accelerated movement of the microparticles forming a massive body, with intensity growing with growth of temperature [7,8]. In experiments [3,6], the weighed samples were heated up to comparatively high temperatures - from ten degrees up to hundreds.

A possible, in such conditions, influence on results of measurements of the thermal air convection, the change of temperature of the scales mechanism, the thermal change of residual magnetization and adsorption of moisture on the surface of samples, and so on – naturally caused caution and even mistrust in estimations of the obtained results. Meanwhile, the results of weighing the heated metal samples were obtained at high enough levels of an effective signal to noise ratio, with the careful account for the influence of the mentioned factors.

In the described experiment, there was carried out the weighing of samples of PZT-piezoelectric ceramics, whose temperature increased by 1.6 degrees in respect to the normal room temperature (24 $^0$C). In so doing, the influence of temperature factors on accuracy of measurements of weight of samples was reduced to a minimum.

The design of the weighed container is shown in Fig. 1.

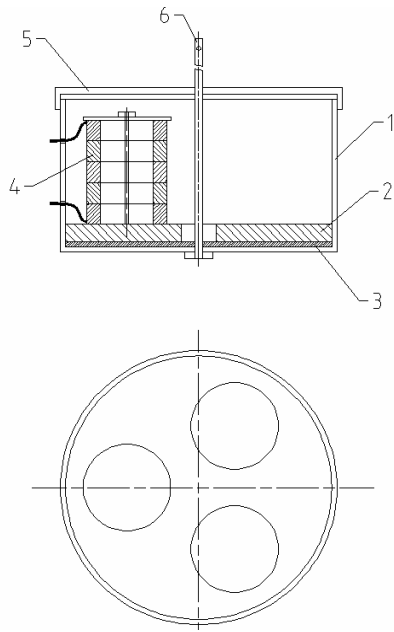

Fig. 1. The arrangement of container. 1- body, 2 – base, 3 – laying, 4 – PZT-pile, 5 – cover, 6 – hanging bar.

The container was placed in the closed box of analytical scales, the high-frequency electric signal was fed to electrodes of piezoelectric ceramics by means of elastic copper conductors 85µm in diameter and 150mm in length. The weighed sample is made in form of three "piles" ("sandwiches") of parallel-connected piezoelectric ceramic rings, 5 rings in each "pile", fixed on the massive brass base; the external diameter of rings is 22 mm, the internal diameter is 16 mm, height is 6 mm; the full weight of 15 rings is equal to 112.9g. In parallel to the power supply terminals of piezoelectric ceramics, there was connected the variable inductance for adjustment of resonance frequency of the supplied signal equal to 389 kHz, which allows to achieve the most effective heating of samples; the amplitude of the resonance signal is equal to 40 V. The readout of scales was carried out by the elongation method with the period of scale beam oscillations equal to 19.7 s. At full weight of the container equal to about 470 g, the error in reading out the changes of weight in time did not exceed 30 mcg.

An example of typical experimental time dependence of the container weight change is shown in Fig. 2.

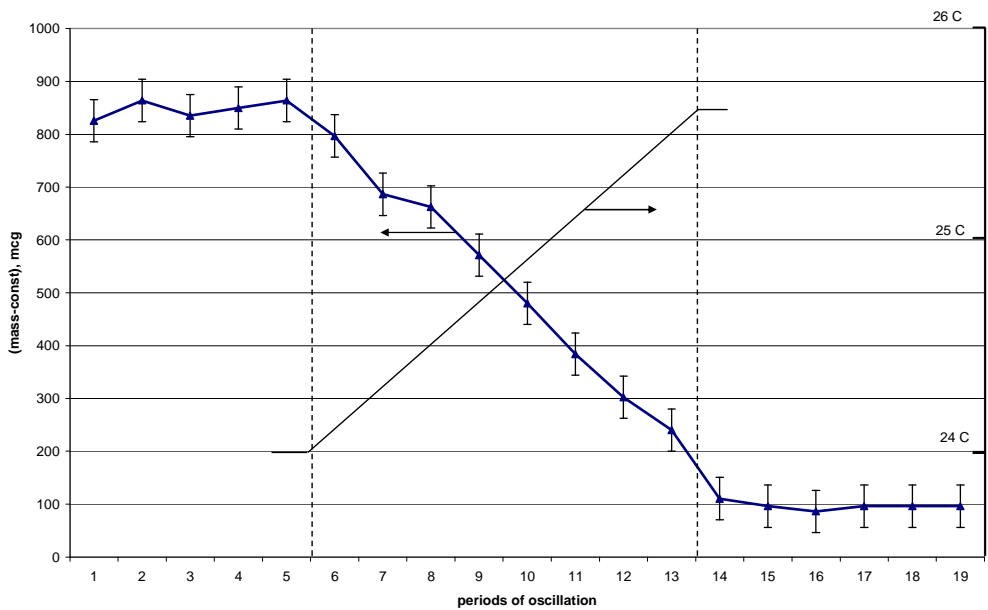

Fig. 2. Experimental time-dependence of container mass by heating PZT-pile from 24.0 till 25.6 $^0$C. Touch lines is "in" and "out" moments. 1 period = 19.7 s.

The change of temperature of piezoceramic sample during the time of heating (2.95 min) is equal to 1.6 $^0$C (it is obtained by control measurements of ceramics temperature before weighing). The temperature of walls of the container remained practically a constant, and short-term heating of air inside the non-hermetical container with volume $V \approx 10^2 cm^3$ even for $\Delta T = 1 \div 2^0 K$ changed the apparent value $\Delta m$ of its weight by no more than 1 mcg ($\Delta m = \rho V \Delta T / T$, where air density $\rho = 1.29 kg/m^3$, $T \approx 297^0 K$).

According to Fig. 2, the relative temperature change $\gamma$ of piezoelectric ceramics weight by 1 degree,

$$\gamma = \left(\frac{\Delta m}{m}\right)\frac{1}{\Delta T},$$

is equal to $\gamma \approx -4.1 \cdot 10^{-6} K^{-1}$.

This value is close to value $\gamma$ for a lead sample, $\gamma \approx -4.6 \cdot 10^{-6} K^{-1}$, obtained in [3].

Let's note that close conformity of $\gamma$ measurement results is realized with essentially different dimensions and configurations of the samples and containers which were used.
So, the laboratory experimental data, obtained in heating of piezoelectric ceramic samples for 1.6 $^0$C, confirm the negative temperature dependence of such sample weights. These data will essentially agree with high-temperature measurements of weight of non-magnetic metal bars [3,6]. The further experimental researches of negative temperature dependence of force of the gravitation, carried out with use of various samples of materials in a wide range of temperatures, will promote the progressive development of physics of gravitation